\documentclass[10pt,a4paper]{report}
\usepackage[latin1]{inputenc}
\usepackage{amsmath}
\usepackage{amsfonts}
\usepackage{amssymb}
\usepackage{graphicx}
\usepackage{epstopdf}

\usepackage[T1]{fontenc}
\usepackage{concrete}
\usepackage{palatino}

\usepackage{a4wide}
\usepackage{amsthm}

\newcounter{lemmas}
\newcounter{definitions}

\newtheorem{lowerbounding_lemma}[lemmas]{Lemma}
\newtheorem{FRM1_lemma}[lemmas]{Lemma}
\newtheorem{FRM2_lemma}[lemmas]{Lemma}
\newtheorem{jsliding_def}[definitions]{Definition}
\newtheorem{jdisjoint_def}[definitions]{Definition}

\newcommand{\FIMUreport}[5]{%
{
\thispagestyle{empty}
\renewcommand{\familydefault}{ppl}
\renewcommand{\encodingdefault}{T1}
\renewcommand{\bfdefault}{bx}

\font \fntfilogo=fi-logo at 3.5cm
\newcommand{\fntfimu}{\fontencoding{T1}\fontfamily{ppl}\fontseries{m}%
     \fontsize{3.5cm}{3.5cm}\selectfont}
\newcommand{\fntstandard}{\fontencoding{T1}\fontfamily{ppl}\fontseries{b}%
     \fontsize{15pt}{17pt}\selectfont}
\newcommand{\fnttitle}{\fontencoding{T1}\fontfamily{ppl}\fontseries{b}%
     \fontsize{23pt}{25pt}\selectfont}
\fntstandard

\noindent
\raisebox{6ex}{\fntfilogo SL}
\hfill
{\fntfimu FI MU}\\[1ex]
\rule[1mm]{1.0\textwidth}{1.2pt}
\begin{flushright}
{\fntstandard Faculty of Informatics\\[.3ex] Masaryk University Brno}
\end{flushright}
\centering
\vfill

\begin{parbox}{\textwidth}{\centering\fnttitle #1}\end{parbox}
\vspace*{1.2cm}

by
\vspace*{1.2cm}

#2
\vfill

FI MU Report Series \hfill  FIMU-RS-#3-#5\\
\raisebox{1ex}{\rule{1.0\textwidth}{1.2pt}}
\raisebox{1ex}{Copyright  
{\leavevmode\setbox0=\hbox{{\fontencoding{OMS}\selectfont\char"0D }}%
 \hbox{\hbox to 0pt{\hbox to \wd0
 {\hss\raisebox{0.0ex}[0mm][0mm]{{\fntstandard c}}\hss}\hss}}%
 \box0%
 } #3, FI MU} \hfill
\raisebox{1ex}{#4\ #3}
\newpage
\thispagestyle{empty}
\newlength{\odsaz}
\settowidth{\odsaz}{Copyright
{\leavevmode\setbox0=\hbox{{\fontencoding{OMS}\selectfont\char"0D }}%
 \hbox{\hbox to 0pt{\hbox to \wd0
 {\hss\raisebox{0.0ex}[0mm][0mm]{{\fntstandard c}}\hss}\hss}}%
 \box0%
 } #3,}
\noindent
\begin{tabbing}
Copyright
{\leavevmode\setbox0=\hbox{{\fontencoding{OMS}\selectfont\char"0D }}%
 \hbox{\hbox to 0pt{\hbox to \wd0
 {\hss\raisebox{0.0ex}[0mm][0mm]{{\fntstandard c}}\hss}\hss}}%
 \box0%
 } #3,\ \= Faculty of Informatics, Masaryk University.\\
\> All rights reserved. \\[1\baselineskip]
\> Reproduction of all or part of this work\\
\> is permitted for educational or research use\\
\> on condition that this copyright notice is\\
\> included in any copy. \\
\end{tabbing}
\vfill

\noindent
\begin{tabbing}
Publications in the FI MU Report Series are in general accessible \\via
WWW: \\[1ex]
\hspace{\odsaz}\ \= {\tt http://www.fi.muni.cz/reports/}\\[4ex]
Further information can be obtained by contacting: \\[1ex]
\> Faculty of Informatics\\
\> Masaryk University\\
\> Botanick 68a\\
\> 602\,00 Brno\\
\> Czech Republic
\end{tabbing}
}}

\title{\bf Employing Subsequence Matching \\in Audio Data Processing}

\author{%
  Petr Voln\'y\\
  xvolny1@fi.muni.cz\\
  \and 
  David Nov\'ak\\
  xnovak8@fi.muni.cz\\
  \and 
  Pavel Zezula\\
  Masaryk University, Brno, Czech Republic \\
  zezula@fi.muni.cz
}

\begin{document}
\FIMUreport%
{Employing Subsequence Matching \\in Audio Data Processing}%
{Petr Voln\'y \\
 David Nov\'ak \\
 Pavel Zezula}
{2011}{August}{04}
 

\setcounter{page}{0}
\maketitle

\begin{abstract}
\noindent
We overview current problems of audio retrieval and time-series subsequence matching. We discuss the usage of subsequence matching approaches in audio data processing, especially in automatic speech recognition (ASR) area and we aim at improving performance of the retrieval process. To overcome the problems known from the time-series area like the occurrence of implementation bias and data bias we present a Subsequence Matching 
Framework as a tool for fast prototyping, building, and testing similarity search subsequence matching applications. The framework is build on top of MESSIF (Metric Similarity Search Implementation Framework) and thus the subsequence matching algorithms can exploit advanced similarity indexes in order to significantly increase their query processing performance. To prove our concept we provide a design of query-by-example spoken term detection type of application with the usage of phonetic posteriograms and subsequence matching approach.
\end{abstract}

\section{Introduction}
In a past couple of decades we have witnessed an enormous rise of the amount of digital data. First, it is caused by the digitization of the data from many branches of human's activity and the sharing of possibly any knowledge today is realized through a digital channels. Furthermore, and especially in recent years in the age of blogs, social networks and services like YouTube, people produce vast amount of digital content. Machines themselves are big content creators too. Let's just mention modern CT scanners or mesh nets of small devices producing constant flow of data about a measured phenomenon.\par
One of the data domains that people have accepted to handle in the digital form is audio. Every moment, we can track a vast amount of new audio data being created. Radio stations, television broadcasting, podcasts, lectures, voice chats -- these all are the instruments of the creation of digitalized audio data and, more specifically, digitalized spoken utterances.
The problem of an Automatic Speech Recognition (ASR) and analysis is more than three decades old. Many approaches were developed to cope with the speech analysis sub-problems like speaker verification, speech transcription, spoken term detection and many others. Solutions suitable for these problems were often related to signal/time-series processing. Modern techniques, like the large vocabulary continuous speech recognition (LVCSR) use large orthographically transcribed speech data to train their sophisticated acoustic and language statistical models and recognizers to achieve good results in automatic speech recognition. The problem can be seen as a classification of a speech where parts of the speech are segmented and classified into known classes, i.e. words from vocabulary. It is expected that subsequent data-mining tasks will be performed on the result of LVCSR.\par 
It is not always feasible to construct such data sets due to the time and expense associated with the annotation of large quantities of audio. This is typically the case for low resource languages for which performing LVCSR is economically less profitable. For example, due to the lack of automatic transcription tools, only a few percent of the recordings of 4,000+ endangered languages, currently being made by linguists, can be analyzed \cite{Boves2009}. It also fails when some out-of-vocabulary words are present in an utterance.\par
Possible drawbacks of LVCSR can be overcome by other approaches like unsupervised methods. Such approaches expects no (or only a minimum) knowledge about the examined utterance and its language specifics and they can be used for pattern discovery in speech, word discovery, or the whole key-phrase detection. It was argued in the literature \cite{Bosch2007, Park2008} that the unsupervised way of analyzing speech, compared with the LVCSR acoustic/language models approach, is much closer to the way of human speech perception and elaboration. The problem is that without a large preprocessing that is present in the LVCSR techniques, the unsupervised methods are very computational expensive because often the traditional similarity distance functions like DTW are extensively employed in the process. This is the problem that we would like to address in our work. We aim mainly on the performance improvements that could be achieved by joining our previous achievements in indexing, similarity searching and subsequence matching areas and one of the state of the art ASR methods. Especially when talking about unsupervised methods, which are much more similar to the classical time series data-mining tasks, we can find many places where our knowledge could enhance the performance of the possible applications. To demonstrate this, we have decided to implement demo application that would employ our general Subsequence Matching Framework, which we are developing, together with one of the known ASR methods and show that many sub-problems of a general ASR can be solved with the subsequence matching approaches.\par
While time series and subsequence matching are natural part in music or general audio retrieval process, we believe that we can employ our framework there also.\par
The rest of this report is organized as follows. In Section 2 we overview an audio retrieval problems form the broader perspective, including music and general audio retrieval. The Section 3 describes the state of the art time series and signal similarity search approaches. In Section 4 we discuss the subsequence matching problem and its applicability for enhancing time series retrieval. In Section 5, our Subsequence Matching Framework and its benefits is introduced and finally the description of our demo query-by-example spoken term detection application based on the framework is presented. The whole report is finished with conclusion and future work directions in Section 6.

\newpage

\section{Audio Retrieval Overview}
Audio retrieval is a very broad discipline ranging from general sound matching over music retrieval to sophisticated speech related methods like speaker identification and automatic speech recognition. In this chapter we make a brief overview of the approaches and techniques for audio related problems that are recognized by the majority of the research community and where, as we believe, it could be helpful to employ advanced techniques for indexing, time series similarity, and subsequence matching. 

\subsection{General Sound Similarity}
In the case of general sound similarity, we have no prior knowledge about the sound like what is the origin of the sound. It must be presumed that it could be emitted by anything from singing birds or see waves to digital sound processor, so one cannot even tell whether the sound is of artificial or natural origin. Because of that, we can use only very low-level features for the sound description. MPEG-7 multimedia description framework \cite{Salembier2002} is one that is considered as a standard for describing multimedia content and it also includes the basic set of seventeen low level audio descriptors for audio features which can be divided into these six groups:
\begin{itemize}
\item basic descriptors -- instantaneous waveform and power values,
\item basic spectral descriptors -- log-frequency power spectrum and spectral features (for e.g. spectral centroid, spectral spread, spectral flatness),
\item basic signal parameters -- fundamental frequency and harmonicity of signals,
\item spectral timbral descriptors -- log attack time and temporal centroid,
\item temporal timbral descriptors -- log attack time and temporal centroid,
\item spectral basis representations -- a number of features used in conjunction for sound recognition for projections into a low-dimensional space.
\end{itemize}
Depending on specific application objectives, these descriptors can be combined in various ways.

\begin{figure}[]
\centering
\includegraphics[]{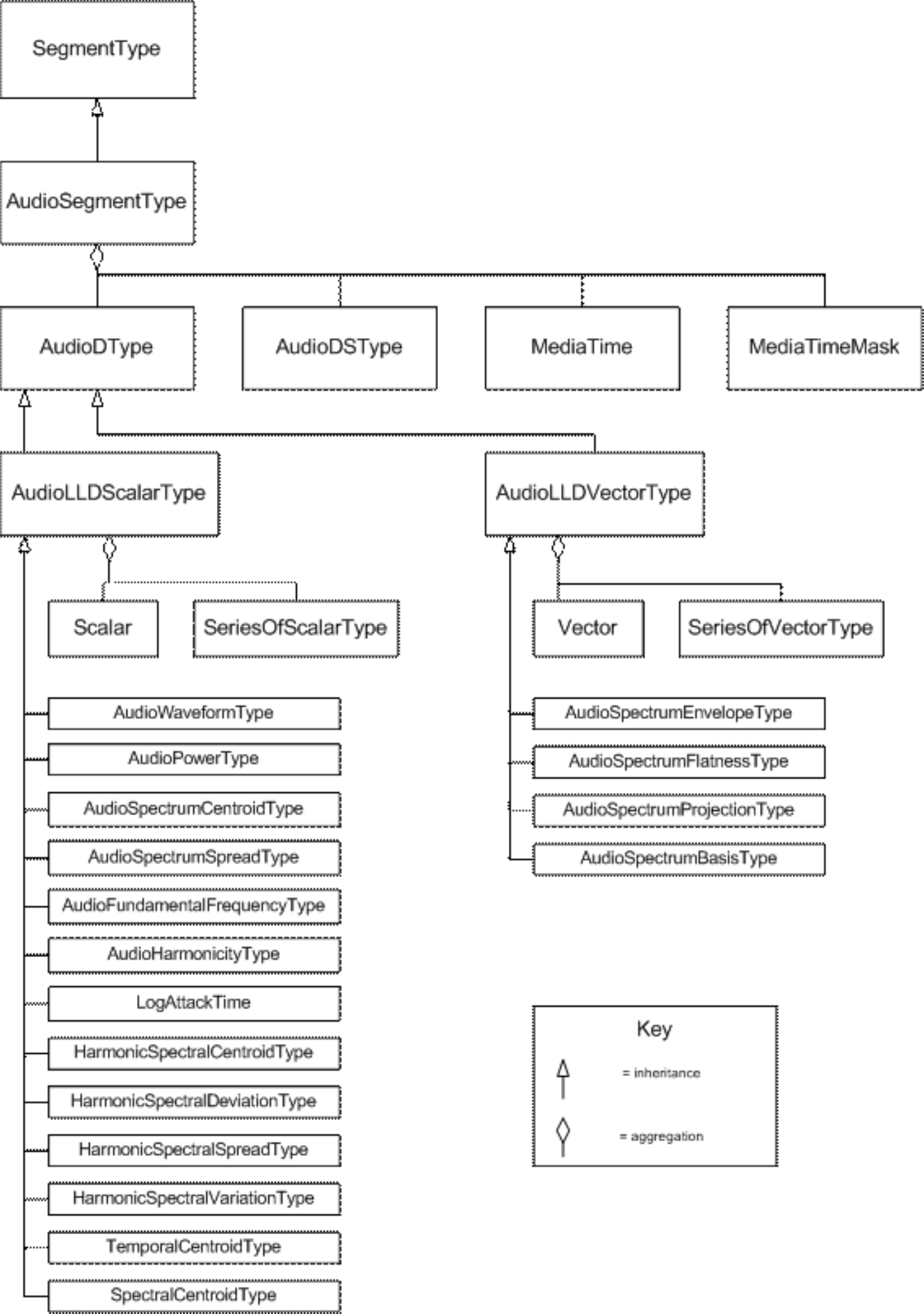}
\caption{Hierarchy of MPEG-7 audio descrirptors. The structure depicts inner classes of an MPEG-7 framework.}
\label{fig:mpeg7_lld_audio}
\end{figure}

\subsection{Music Information Retrieval}
Music information retrieval, herein referred to as MIR, covers a broad range of topics. Although, the idea of using computers for the purpose of MIR is about four decades old, the true explosion of interest in this topic took place in 1990's. One of the factors that contributed to it was an establishment of MIDI format as basis for music information sharing. MIDI, as clearly-defined mathematical-based musical format, simplified music analysis and processing and brought new possibilities for music indexing. The next milestone was wide utilization of compression formats such as MP3 and Ogg Vorbis together with the increasing amount of freely available musical databases available for searching. Particular approaches for MIR are partly derived from the representation of the music that one works with; therefore we make a short overview of these.

\paragraph{Sampled Audio}
Very accurate when it comes to interpreting the music but very expensive from the MIR point of view. The reason for this is that it consists of thousands of samples of sound taken every second resulting in potentially very large files. Furthermore, sampled audio representation is not robust against noise or small variations of music so it can eventually decrease the performance of the query. Therefore, sampled audio representation is rarely used as an audio description format.

\paragraph{Musical Instrument Digital Interface}
MIDI is a comprehensive type of a representation. It means that it is suitable for quite accurate recreating of an original music data. In contrast to sampled audio, it is much less verbose format. Instead of producing thousands samples per second depicting frequency and amplitude of the sound, it works with discrete alphabet of tones and instruments and it tries to capture when the particular tone starts and where it ends. Although, the re-created sound cannot reach the quality of sampled audio, it is accurate enough for the listener to recognize the original tune. Moreover it is much more robust against background noise and it is much cheaper to process.

\paragraph{Reduced Representations}
The goal of members of this class of representations is not to transcribe complete musical information but only some of its features. It is not usually possible to reconstruct the original music but it gives us good amount of information about features on which basis we can quite easily compare two pieces of an audio data. Here, again, we would like to mention MPEG-7 standards and its audio descriptors. These can be expressible by sole number - scalar descriptors, or holding more complex information in a form of multidimensional vectors. MPEG-7 descriptors and theirs membership to one of the mentioned groups can be seen on Fig. \ref{fig:mpeg7_lld_audio}.

\subsubsection{MIR Applications}
There are several types of applications that employ MIR and they differ in a purpose of usage. One big class is a query by humming applications where an application expects human voice as an input which can be either a bunch of tones that should form a melody or even a part of the song hummed with lyrics. Other type of such an application could be genre detection or instrument detection; there are also applications that try to guess the mood of the tune and so on. Each needs a bit different set of features to be extracted or comparing them in a bit different way.

\paragraph{Query by Humming}
Query by humming usually referrers to a collection of input methods where the human voice is required to produce the query. The method was popularized by Ghias et. al in 1995 \cite{Ghias1995} and many works tried to enhance the query by humming approach \cite{Zhu2003a} and sometimes it was also with the usage of MPEG-7 descriptors \cite{Batke2004}. Nowadays, the technology in indexing and querying vast musical databases allows commercial projects to offer query by humming type of application to the widest audience. As an example lets mention SoundHound; it offers mobile application where user hums a small part of tune (maybe he wants to know the name of the song or want to buy it) and gets full name of the song and other metadata with the immediate possibility to buy it.

\subsection{Speech Related Areas}
For humans, speech is the most natural form of communication, which leads to the fact that speech processing is one of the most exciting areas of the signal processing. In spite of the great effort of the research community, there is still significant communication gap between humans and machines. The spoken word would be the perfect interface for many real-life applications. Although we can observe rising number of applications that uses human voice as an input, solving speech processing problems is still very challenging.\par
\begin{figure}[]
\centering
\includegraphics[scale=0.75]{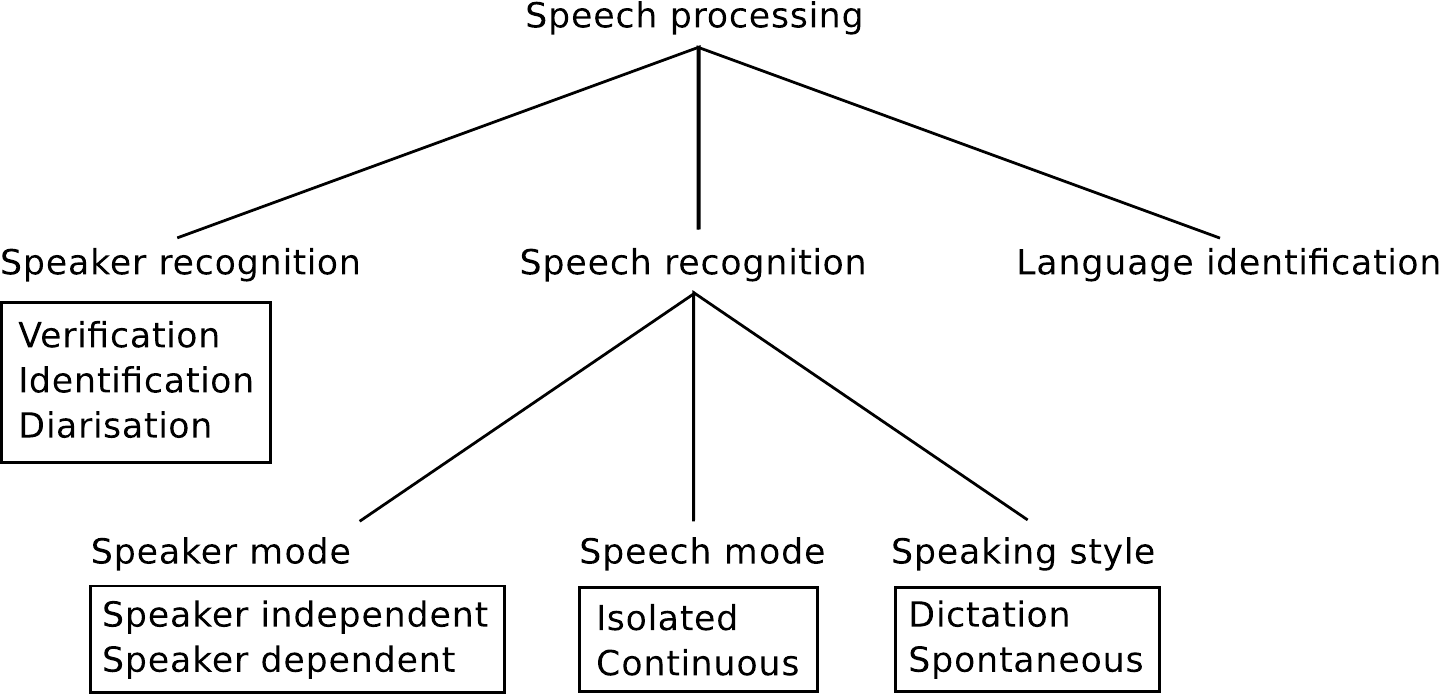}
\caption{Overview of the speech processing areas.}
\label{fig:speech_apps}
\end{figure}

In this section we make a brief overview of the most significant areas related to speech, which are depicted on Fig. \ref{fig:speech_apps}. We name three basic application classes: speaker recognition, speech recognition, language identification. Speaker recognition can be further divided into speaker verification, identification and diarisation. Speaker verification, text dependent or text independent, is the case where we want to verify the identity of the person through specifics of his voice. That means that the system knows which person should be verified and the output is either yes or no or maybe some probability.\par
On the other hand, in the case of speaker identification, the system tries to guess who is speaking. In other words, there is a database of speakers and an application matches an input voice with indexed entries. Speaker diarisation is a method for distinguishing between speakers in the multi-person conversation and it tries to divide the conversation into utterances that belong to particular speaker.\par
There are also very specific areas like gender recognition for recognizing a gender of a speaker or language identification. The recognition method is usually based on statistical models of male/female speakers and models for specific languages.\par
Finally, there is a wide area of speech recognition applications, where one can find many variables for the particular application like whether it is speaker dependent or speaker independent, whether it needs to perform the recognition for continuous speech or preprocessed short utterances, or whether the speaker is aware that his words are to be recognized and she sort of dictates the text or if the system should recognize any spontaneous speaking style. All these variables influence the choice of the proper ASR approach for the specific application.

\subsubsection{Automatic Speech Recognition}
One of the most popular approaches for the ASR is the large vocabulary continuous speech recognition (LVSCR). This technique usually uses specific models for each language and a decoder that extracts posterior probabilities of sub-word phonetic units. With the aid of the trained models (usually Hidden Markov Models) the input in the form of sub-word phonemes is classified as one of the words in vocabulary. These sub-words phonetic units are recognized by so-called phonetic recognizers. In this phase the acoustic model of the language is used to produce posterior probabilities of the sub-word units for the raw audio input. The output can be represented as a confusion network. In the case that we also have the knowledge about the speaker of the analyzed utterance, we can also employ a speaker specific acoustic model that was trained on the sample data produced by that speaker.\par
Basic components of a speech recognition process are front end and decoder. Decoder usually can not work with raw audio data. It works only with features extracted by front end. A good example of a wide spread representation is Mel-Frequncey Cepstral Coefficients (MFCC) \cite{Wei2006}. MFCC is the most widely used among acoustic signal representations \cite{Rabiner1993, Picone1993} and not only for the sake of speech recognition (it is widely used for a speaker recognition tasks too).\par
\begin{figure}[]
\centering
\includegraphics[scale=0.75]{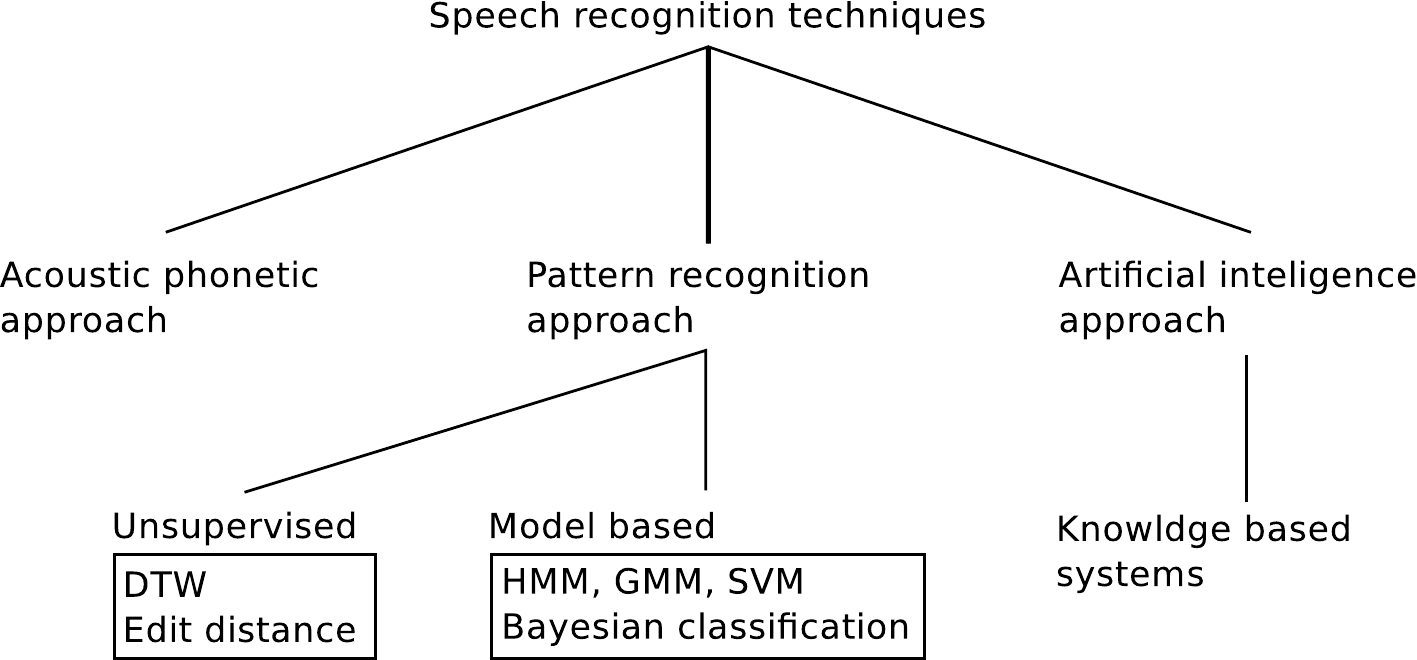}
\caption{Taxonomy of techniques used in ASR.}
\label{fig:speech_techniques}
\end{figure}

The taxonomy of speech recognition related techniques can be seen on Fig. \ref{fig:speech_techniques}.

\subsubsection{Unsupervised Methods}
Unlike the model based approaches, the pattern recognition based on unsupervised methods have no, or only small, prior knowledge about the language being used or the speaker. In other words, it is possible to use this approach even for languages or dialects where no linguistic corpora are available. The unsupervised methods \cite{Huijbregts2011, Bosch2007, Park2008, VonZeddelmann2010} represent quite recent stream of research and it can be seen as a `back to the roots' approach because it has much more common with the early ASR methods than with the current state of the art model based ways of ASR.\par
 Unsupervised methods represent departure from traditional models of speech recognition, where the goal is to classify speech into categories defined by a given inventory of lexical units -- phonemes or words. Instead, such an inventory is discovered in an unsupervised manner, where words, sub-words, word-like units or generally patterns in speech are extracted by exploiting the structure of repeating patterns within untranscribed audio stream.\par
As we observed, methods used for unsupervised extraction of such a patterns are often modifications of linear programming algorithms based on DTW approach. This might be one of the causes for performance issues of the unsupervised approaches and the reason for the fact that current state of unsupervised techniques do not allow to achieve the same performance as a model based approaches.

\newpage

\section{Time Series Processing: State of the Art}
Time series data mining tasks got a lot of attention in last decade. It was caused by the data explosion in many branches and the need to handle vast amounts of data effectively and efficiently. In this section, we provide the reader with an introduction to the mentioned topics. We overview the most important similarity functions and data representations that are meaningful from our point of view. Furthermore, we discuss recent problems related to the outlined areas. Generally, there are four different tasks in time series data mining research \cite{Keogh2002}:
\begin{itemize}
\item Indexing (Query by content): Given a query $Q$, similarity function $D(Q,S)$ and indexed data in DB, find the nearest match in DB.
\item Clustering: Find natural groupings of the time series in database DB under some similarity/dissimilarity measure $D(Q,S)$.
\item Classification: Given an unlabeled time series $Q$, assign it to one of two or more predefined classes.
\item Segmentation: Given a time series $Q$ containing $n$ data points, construct a model $Q$ , from $K$ piecewise segments
$(K << n)$ such that $Q$ closely approximates $Q$.
\end{itemize}
We aim mostly at the first case, but you will see that some approaches can be used to solve more than one of the mentioned tasks. First, we explain the lower bounding lemma and why it is so important.

\subsection{Lower Bounding}
The lower bounding lemma was introduced in \cite{Faloutsos1994} and it is very desirable property of representations and similarity models. It enables to prune big parts of the space being searched in a computationally cheap way and it guarantees no false dismissals in an answer (but it can contain false alarms). The key advantage of this approach is that we use the expensive true distance function only on the small subset of the indexed data which were selected as the answer candidates by the cheaper lower bounding mechanism.

\begin{lowerbounding_lemma}
To guarantee no false dismissals for range queries, the extraction function $F$ and distance functions $Dist$ and $LB\_Dist$ should satisfy the following formula:
\end{lowerbounding_lemma}
$$ LB\_Dist(F(S), F(Q)) \leq Dist(S, Q) $$

For instance, it was proven \cite{Faloutsos1994} that DFT is lower bounding transformation for Euclidena Distance.\par
To compare representations according to their lower bounding properties the tightness of lowerbound measure \cite{Dinga, Keogh2000} is used. Its vale is between 0 and 1 inclusive and is computed as:
$$ TLB = LowerBoundDist(S ,Q) / TrueEuclideanDist(S, Q)$$
The higher the $TLB$ value, the better the lower bound. $TLB$ seems to be very meaningful measure and it seems to be the current consensus in the literature \cite{Cai2004, Chen2004, Chen2005a, Keogh2006}. The advantage of the measure is that it is implementation-free and independent on external variables, e.g. software or hardware testing infrastructure. With this measure one can easily predict indexing performance.

\subsection{Representations}

\begin{figure}[]
\centering
\includegraphics[scale=0.75]{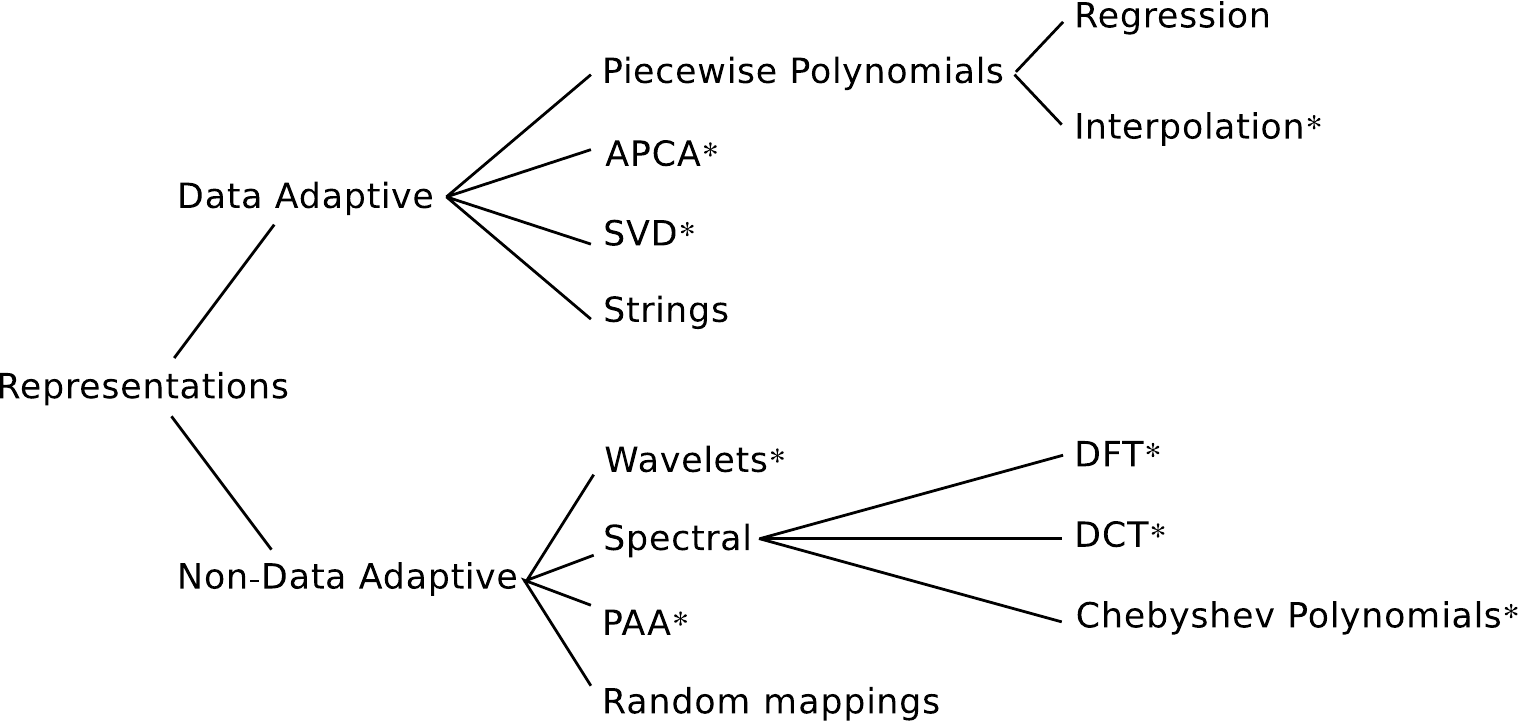}
\caption{Basic hierarchy of representations. Those with asterisk after the name allows lower bounding.}
\label{fig:representations}
\end{figure}

Dimensionality reduction is one of the most important tasks in time series processing. The desirable features of the transformed representation are a good ratio between the level of reduction and the accuracy of an approximation of the original time series data and the ability to provide lower bounding options for a distance function. In this section we provide a brief overview of meaningful representations that were widely accepted by the research community (See Fig. \ref{fig:representations}).\par
Generally, we can look at representations from several points of view:
\begin{itemize}
\item data adaptive/non-data adaptive -- the difference is whether the method reflects the features of the time series being transformed,
\item continues/discrete domain representations -- coefficients are from some continuous domain, e.g. real numbers, or from finite set of values (string representations) symbols; The latter is often called string representation,
\item allows/disallows lower bounding -- is it proven, that we can lower bound true distance on the original time series with the reduced representation?

\end{itemize}
The basic division of representations that we are going to overview can be seen in Fig. \ref{fig:representations}.
\subsubsection{Non-data Adaptive Representations}
\paragraph{Discrete Fourier Transform}
Discrete Fourier Transform was the first proposed method for time-series dimensionality reduction \cite{Faloutsos1994}. It transforms a time-series from the time/space domain into the frequency domain. A fast algorithm exists for such a transformation in $\Theta (n\log n)$ time \cite{Cooley1965}.
The DFT represents the time-series data by using Fourier coefficients (a combination of sine and/or cosine waves, each represented as a complex number). Only the first few coefficients are required to approximate the original time-series since the slow changing/low frequency part of the sequence is the underlying shape of the sequence. The coefficients beyond the first few have such low amplitude that they only provide a very small gain in representative accuracy but produce a greater decrease in the compression.
An advantage of DFT is that it maintains the property of allowing Euclidean distance calculations. Due to the exclusion of some coefficients, and therefore some positive values, the Euclidean distance calculation of the two DFT's guarantees the lower bound of the Euclidean Distance \cite{Faloutsos1994}.
\paragraph{Discrete Cosine Transform}
A discrete cosine transform (DCT) \cite{Korn1997} expresses a sequence of finitely many data points in terms of a sum of cosine functions oscillating at different frequencies. In particular, a DCT is similar to the discrete Fourier transform (DFT), but using only real numbers. DCT is  operating on real data with even symmetry (since the Fourier transform of a real and even function is real and even), where in some variants the input and/or output data are shifted by half a sample.\par
DCT also takes place in a process of extracting more sophisticated audio descriptors like Mel Frequency Cepstrum Coefficients (MFCC).
\paragraph{Discrete Wavelet Transform}
Discrete Wavelet Transform (DWT) \cite{Chan1999} is similar to DFT in many aspects, except that rather than using sine and cosine functions, it uses wavelets as base function. Wavelets are basis functions used in representing data or other functions. Wavelet algorithms process data at different scales or resolutions in contrast with DFT where only frequency components are considered.\par
Another difference is the result of the transform. While the DFT transform the time-series into the frequency domain, DWT transforms the time-series into the time/frequency or space/frequency domain.
Haar wavelets \cite{Chan1999} are often used as s good example of DWT approach because of its good pruning power (i.e it is a good lower bound of Euclidean distance).
\paragraph{Piecewise Aggregate Approximation}
According to Piecewise Aggregate Approximation (PAA), to reduce the data from $n$ dimensions to $N$ dimensions, the data is divided into $N$ equi-sized ``frames''. The mean value of the data falling within a frame is calculated and a vector of these values becomes the data reduced representation \cite{Keogh2000}.

\subsubsection{Data Adaptive Representations}

\paragraph{Single Value Decomposition}
Singular Value Decomposition (SVD) \cite{ Korn1997, RaviKanth1998} is an optimal dimensionality reduction technique on matrices under the Euclidean metric. The SVD reduction is suitable for use on any matrix, including applications to indexing images and other multimedia objects. Keogh et al. \cite{ Keogh2000} first used the SVD on time-series data, noting its advantages and disadvantages. The difference with the other proposed techniques as opposed to this one is that the others are local transformations, but SVD is a global transformation.
\paragraph{Adaptive Piecewise Aggregate Approximation}
APCA is a modification of PAA that allows arbitrary length frames in contrast with PAA that divides a time-series into $N$ frame of equal length. This allows indexing using standard multi-dimensional index structures by mapping each mean value to each dimension.
APCA differs from PAA in that by allowing arbitrary length frames, an extra value must be stored to identify the length of each window. This led researchers to believe APCA did not allow indexing \cite{ Yi2000}. Chakrabarti et al. \cite{ Chakrabarti2002} showed that APCA could indeed be indexed using multi-dimensional indexing structures. An APCA representation requires two numbers per window.


\subsection{Distance Measures}
\begin{figure}[htbp]
\centering
\includegraphics[scale=0.75]{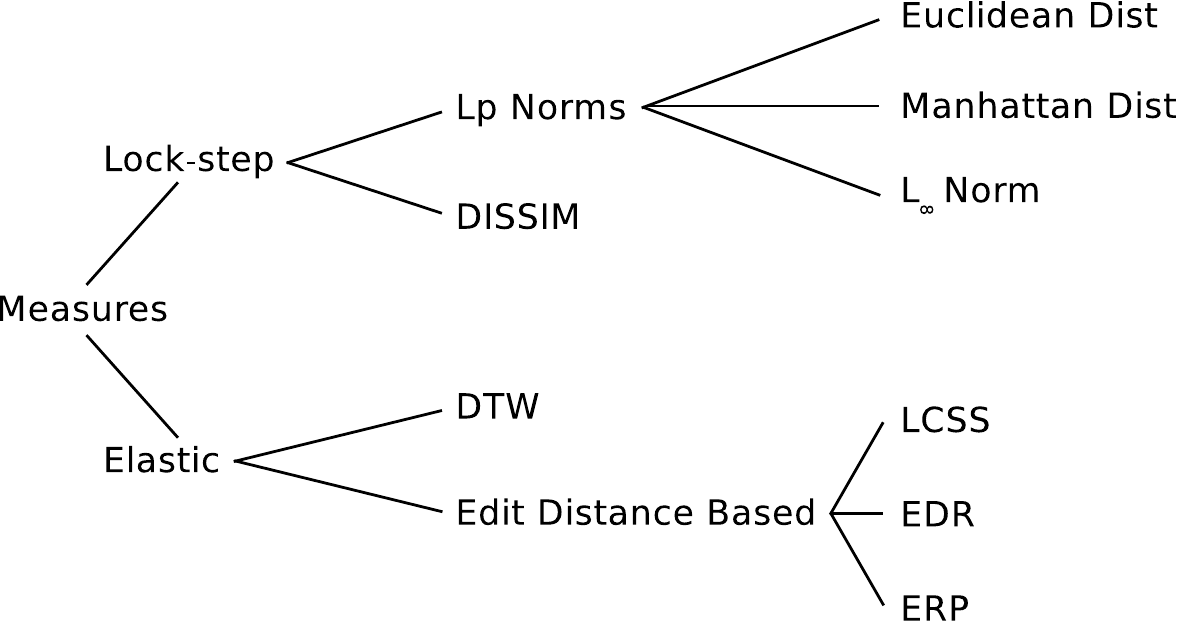}
\caption{Basic hierarchy of distance functions.}
\label{fig:measures}
\end{figure}
Measuring the distance between objects and thus the similarity is essential for any data mining task. In a perfect case, the distance measure would be cheap to compute, invariant against all possible transformations and with the possibility to parametrize it to give results that are close to  human perception of similarity on the given objects domain. Futhermore, it is desireable for the function to obey axioms of a metric space, because when those axioms are satisfied, we can use advanced indexing techniques and perform queries on the indexed data very effectively \cite{Zezula2005}.\par
Metric space $ \mathcal{M}=(\mathcal{D},d) $ is defined as a set of objects $\mathcal{D}$ and the distances between them. The distance is computed by the function $d:\mathcal{D}\times\mathcal{D}\rightarrow\mathbb{R}$ that must satisfy the following porperties. $\forall x,y,z\in\mathcal{D}$ hold:
\begin{itemize}
\item $ d(x,y)=0\Leftrightarrow x=y $ (identity)
\item $ d(x,y)=d(y,x) $ (symetry)
\item $ d(x,z)\leq d(x,y)+d(y,z) $ (triangle inequality)
\end{itemize}
The key axiom is triangle inequality which can be exploited to prune big parts of the indexed space and return the best matching objects very quickly. Common types of queries is a range query and kNN query. First one returns a set of objects that are within the given diameter $r$ from the query object $q$. The latter one returns $k$ nearest objects to the query object $q$.

\subsubsection{Lock-step Measures}
This class of distance functions is suitable only for time-series with the same length. It measures the distance between every particular part of the sequence and expects that these parts were taken in the same time interval.

\paragraph{Lp Norms}
Lp norms is a set of functions that assign strictly positive value to a vector in a Lp space. We will consider only two special cases of Lp norms. The ubiquitous Euclidean norm and Manhattan norm. \par
Although very simple and straightforward, Euclidean distance is one of the most used similarity measures in time series data mining. It is a classical lock-step similarity method so it can be used only for comparing segments of the same length. It is not robust against time shifting, scaling  and to any other transformations. As we said the Euclidean norm is a special case of Lp norm where $p=2$, so the length of vector $x=(x_1,...,x_n)$ is defined as:
$$ \left|\left|x\right|\right| = \sqrt{ \sum_{i=1}^{n} {\left|x_i\right|}^2} $$
It has been shown that even with the outlined disadvantages of Euclidean distance function it is very reasonable to it use in many fields of time series data mining. Good feature of this measure is that it satisfies properties of a metric function.\par
The second Lp norm we are about to mention is a Taxicab norm or Manhattan norm. In this case the $p=1$.

\begin{figure}[htbp]
\centering
\includegraphics[scale=0.75]{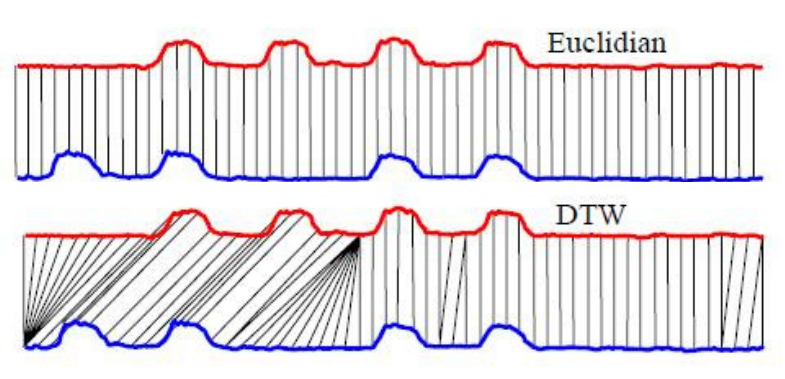}
\caption{Euclidean distance vs. DTW.}
\label{fig:eucl_dtw}
\end{figure}

\subsubsection{Elastic Measures}
These measures allows to compare sequence even if they have different lengths. It is allowed by the ability of time warping (See Fig. \ref{fig:eucl_dtw}).

\paragraph{Dynamic Time Warping}
Dynamic Time Warping (DTW) ads sort of elasticity to the process of the time series comparison. It allows warping of sequences in a time to eliminate scaling or gaps on the time axis. Semantically, it means that for example in an audio retrieval process, we can identify a spoken word even if it was spoken in a different tempo than is the tempo of the indexed reference word. On Fig. \ref{fig:eucl_dtw} you can see the difference in a way of pairing values of the series being compared by fixed step Euclidean Distance and elastic DTW. Generally, DTW tries to find an optimal match between two sequences. More formally, Dynamic Time Warping is a linear programming method for finding a minimum cost path in an accumulation matrix.\par
The problem of DTW is a performance and sometimes inappropriate behavior, when it applies a lot of warping. The performance problem of DTW is also given by the fact, that it is not a metric function and thus cannot be indexed as a metric space.\par
\begin{figure}[htbp]
\centering
\includegraphics[scale=0.35]{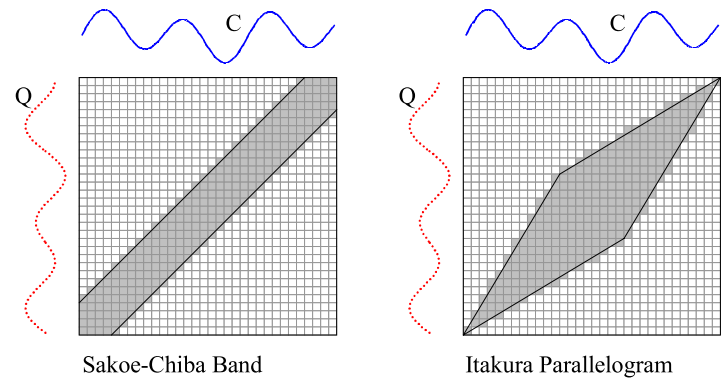}
\caption{Global constraints limit the scope of the warping path, restricting them to the gray areas. The two most common constraints in the literature are the Sakoe-Chiba Band and the Itakura Parallelogram \cite{Keogh2004}.}
\label{fig:dtw_constraints}
\end{figure}

The problem of performance and misbehavior was addressed by the introduction of Sakoe-Chiba band \cite{Sakoe1978} and Itakura Parallelogram \cite{Rabiner1993} which ideas were to constraint the search space only to a part of the matrix along the diagonal path (See Fig. \ref{fig:dtw_constraints}).\par

To enhance the performance of the data retrieval applications that were using the DTW method, the lower bounding functions for the DTW that would be indexable were proposed. The lower bounding function introduced by Kim et al. \cite{Kim2001} (known as LB$\_$Kim) extracts 4 numbers as a feature vector for each sequence. The features are the first and last elements and the minimum and maximum values.\par
Another lower bounding function LB$\_$Yi was introduced by Yi et al. \cite{Yi1998} and it exploits the fact that all points in one sequence that are larger (smaller) than the maximum (minimum) must contribute at least squared difference of their value and the maximum (minimum) value of the other sequence in the final DTW distance.\par
Although, they both added the desirable property of lower bounding, the lower bound was still too loose. This issue was addressed by Keogh et al. in the work \cite{Keogh2004} introducing LB$\_$Keogh, which idea is to construct envelopes (based on the idea of global constraint of Itakura parallelogram) around the original signal and use PAA representation for both \textit{Upper} and \textit{Lower} envelope borders. On these reduced envelope representations the special lower bounding function LB$\_$PAA (computes the ED between the envelopes) is computed, and Minimum Bounding Rectangles (MBR) covering the original sequence are created.  Then with the aid of $MINDIST(Q,R)$ function, where $Q$ is the query sequence and R is the MBR, the lower bounding distance is computed.\par
To our best knowledge, LB$\_$Keogh is the best lower bounding technique for DTW approaches.

\paragraph{Edit Distance Based Functions}
To overcome the fact that DTW, in its pure form, is not eligible for exact indexing, i.e. indexing in metric spaces, the inspiration from edit distance based approaches gave a birth to the Edit Distance on Real Sequence (EDR) \cite{Chen2005} and Edit Distance with Real Penalty (ERP) \cite{Chen2004}. They both exploit the observation that the warping in one sequence can be seen as a gap addition in the other sequence, which takes the problem of warping to the level of edit distance like problem. The main difference between the DTW approach and edit distance based approaches is the penalization for warping which the DTW approach lacks.

\newpage
\section{Subsequence Matching strategies}
So far have we discussed representations and similarity models that are used for general time series data mining tasks but we have not mentioned anything about subsequence matching yet. In general, subsequence matching is a specific problem in the area of sequence matching. Therefore, all the outlined methods are substantial to both sequence matching and subsequence matching. The latter differs in indexing and retrieval strategies but somewhere in the process it goes down to one of the mentioned similarity measuring methods on the given representations.\par
A specific application is heavily determined by its data meaning and interpretation. So prior to the inventing subsequence matching strategies one must ask few substantial questions. What is the meaning of the data? Do we seek for trends or patterns? Is it possible to compare data with different lengths? Is time warping a desirable feature during search? What will be the size of the window? What will be the volume of the indexed data? etc. As you can see the process has many variables that we must take into account before we start tailoring the subsequence matching approach for the specific problem.\par
This section covers basic problems and techniques used in subsequence matching applications. We will discuss usage of sliding and disjoint windows during indexing and querying, the effect of the chosen windows size on the performance of the application, discussion on where to employ dynamic time warping and where to use Euclidean distance, and finally specific problem of finding motifs in the time series.\par
In the following text we will use this notation:

\hspace{1cm}
\begin{center}
\begin{tabular}{|l|l|}
\hline
Symbols & Definitions \\
\hline
\hline
$Len[S]$ & The length of sequence $S$ \\
\hline
$S[k]$ & The k-th value of the sequence $S$ \\
\hline
$S[i:j]$ & Subsequence of $S$ including values between $S[i]$ to $S[j]$ inclusive. \\
\hline
$D(Q,S)$ & Distance between two sequences $Q$ and $S$ \\
\hline
$s_i$ & The $i$-th disjoint window of sequence S \\
\hline
$\omega$ & Length of the (sliding/disjoint) window \\
\hline
$\epsilon$ & User defined tolerance \\
\hline
\end{tabular}
\end{center}

\hspace{1cm}

\subsection{Sliding And Disjoint Windows}
For the sake of subsequence matching, we need to extract subsequences both when indexing and querying data. There are two related major techniques for creating subsequences:
\begin{itemize}
\item disjoint window -- it divides the given sequence into disjoint windows. where one starts where the other ends,
\item sliding window -- it creates all possible windows of given length $\omega$ that can be extracted from the given series. In other words, if one starts at position $i$ within the sequence, the other start at position $i+1$.
\end{itemize}
Both, sliding and disjoint windows are used in subsequence matching techniques to achieve the state, where each window created from the query can be compared to each window in an index. First time, this concept was used in field opening paper \cite{Faloutsos1994} by Faloutsos et al. (FRM in short). There, they use sliding windows for indexed data and disjoint windows for queries. To reduce the amount of subsequence to subsequence distance computations, not all subsequences are indexed in the database, but minimum bounding rectangles representing a bunch of windows (MBR) are used instead. So in the case the range query is initiated, query disjoint windows are first compared to the indexed MBRs and only those MBRs, where at least part of MBR is in the range, are selected as a candidates for the answer. Then all disjoint windows from the candidate MBRs are taken and all false alarms (windows not in range) are found and dismissed from the precise answer. Although, this lower bounding technique ensures no false dismissals, it is not very efficient and produces a lot of false alarms.\par
The opposite way of using sliding and disjoint windows was introduced by Moon et al. \cite{Moon2001}. They divide data sequences into disjoint windows and the query sequences into sliding windows. Hence, this approached exploits duality in constructing windows, it was called DualMatch. It has been shown that the duality based approach is correct (i.e. it incurs no false dismissals) and that it outperforms FRM. DualMatch reduced drastically the number of points that need to be stored to $1/\omega$ of that of FRM. The disadvantage of DualMatch is the upper bound for window size which in return caused additional false alarms due to the \textit{window size effect} (see section 5.2)\par
Another method, GeneralMatch, was brought by Moon et al. in 2002 \cite{Moon2002}. It is based on a generalization of constructing windows. The authors introduced the concept of $J$-sliding and $J$-disjoint windows  with the specified $J$ sliding factor and they defined them as follows:

\begin{jsliding_def}
A $J$-sliding window $(1 < J < \omega)$ $s_{i}^J$ of size $\omega$ of the sequence $S$ is defined as the subsequence of length $\omega$ starting from $S[(i-1)*J+1](1 < i < \frac{Len(S)-\omega}{j}+1)$.
\end{jsliding_def}

Intuitively speaking, if we have $\omega=16$ and $J=4$, we construct windows by shifting subsequence of length 16 and by 4 entries, and thus, the starting points of the 4-sliding  windows are $S[1],S[5],S[9],...,$ respectively.

\begin{jdisjoint_def}
A $J$-disjoint window $(1 < J < \omega)$ $q_{(i,j)}^J$ of size $\omega$ of the sequence $Q$ is defined as the subsequence of length $\omega$ starting from $Q[i+(j-1)*\omega](1\leq i\leq J,1\leq j\leq \frac{Len(S)-i+1}{\omega}))$ in $Q$
\end{jdisjoint_def}

Again, if we have an example where $\omega=16$ and $J=4$, we construct windows $Q[i:i+\omega -1], Q[i+\omega : i+2\omega -1],$... by dividing Q[i:Len(Q)] into disjoint windows for every $i$ where $(1<i<4)$.\par

\paragraph{I/O Issues}
Logically, when we divide the sequences into many subsequences, the amount of comparisons during the query processing rises. This also implies that subsequence matching algorithms do lots and lots of I/O operations needed for fetching particular subsequences one by one again and again. One of the advanced subsequence matching methods addressed this problem by introducing \textit{deferred group subsequence retrieval} \cite{Han2007}. This technique tries to cope with the fact that, because of excessive disk I/O operations and bad buffer utilization, we might have to read same data pages repeatedly from the disk. The basic idea of the proposed solution is to delay a fixed size set of subsequence retrieval and enable batch retrieval.


\subsection{Window Size Effect}
The effect of a window size on a performance was first discussed in previously mentioned DualMatch approach \cite{Moon2001}. The windows size effect is caused by the application of two lemmas that were introduced in FRM:

\begin{FRM1_lemma}
When two sequences $S$ and $Q$ of the same length are divided into $p$ windows $s_i$ and $q_i$ $(1 \leq i \leq p)$ respectively, if $S$ and $Q$ are in $\epsilon$-match, then at least one of the pairs $(s_i, q_i)$ are in $\epsilon/\sqrt{p}$-match. That is, the following equation holds:
$$ D(S,Q)\leq\epsilon \Rightarrow \bigvee_{i=1}^p D(s_i,q_i) \leq \epsilon/\sqrt{p}$$
\end{FRM1_lemma}

\begin{FRM2_lemma}
If two sequences $S$ and $Q$ of the same length are in $\epsilon$-match, then any pair of subsequences $(S[i:j], Q[i:j])$ are also in $\epsilon$-match. That is, the following equation holds:
$$ D(S,Q)\leq\epsilon \Rightarrow D(S[i:j],Q[i:j]) \leq \epsilon $$
\end{FRM2_lemma}

Application of Lemmas 2 and 3 for long query sequences causes false alarms. That is, when two sequences $S$ and $Q$ are divided into $p$ windows $s_i$ and $q_i$ $(1 \leq i \leq p)$ respectively, although a pair $(s_i, q_i)$ is in $\epsilon/\sqrt{p}$-match, the distance between $S$ and $Q$ may be greater than $\epsilon$. To reduce this kind of false alarms, it is reasonable to use as large windows as possible. For example, let the window size of the method A be twice as large as that of the method B. Then, by Lemma 2 or 3, a candidate subsequence of the method A must also be a candidate of the method B. However, the inverse does not hold. This effect was defined as the \textit{window size effect} \cite{Moon2001}. The size of the window, however, must be less than or equal to the length of the query sequence; thus, the maximum window size depends on the length of the query sequence. Typically, the algorithm have the problem that the performance decreases as the difference between the query sequence length and the window size increases. This problem was addressed in \cite{Koh2005, Lim2007}. The proposed techniques employ creating multiple indexes and their usage based on the query parameters. The dependency of the performance degradation on the number of queries and query vs. window size was revealed and it was claimed that the need for multiple indexes tailored for the variable query lengths is crucial. Both mentioned articles introduced heuristic methods for determination of the window sizes for particular indexes based on the distribution of the query sequence lengths.

\subsection{When We Need To Warp?}
Euclidean Distance and Dynamic Time Warping based techniques have both many advocates. Euclidean distance is mostly favored for its simplicity, lower bounding options and the fact that it could be easily computed. On the other side stands much more sophisticated linear programming approach of dynamic time warping. During the past few decades the latter pushed the former out of many applications and it was claimed that the dominance of DTW over ED is inevitable. There are areas where this has become true like in speech recognition and query-by-humming where the warping ability is crucial to match the spoken terms with words from the database or where one can expect that the humming will not have the same tempo as the indexed data that we want to match with the query.\par
Despite the fact that DTW outperforms ED in these areas, arguments in the favor of ED were raised in the literature \cite{Ratanamahatana2005, Ann2004}. It was argued that the bigger the data collection, the lesser the need for DTW because it often degrades to the same result as the ED measure. This is caused by the fact that the probability, that there are some data in the collection that are very similar to the possible query, grows with a size of the indexed data collection. So the usage of DTW and ED in this case would bring very similar results but ED is cheaper to compute.\par
On the other hand, this reasoning cannot be applied generally. We believe that for some domains the warping feature of the DTW is crucial even if the collection is large enough. Again, on the example of the spoken term detection, we do not want to find only the one best match, but to find possibly all variations of the spoken term and this is the case where DTW, or other functions with the warping ability, are irreplaceable. In the case that the warping would be needed only because the different offset of the data, than, when considering subsequence matching, it is better to use ED. We do not need DTW in this case because the offset difference of the sequence can be overcome by the sliding/disjoint windows.

\subsection{Finding Motifs In Time Series}
We can say that finding motifs \cite{Lin2002} is specific sub-problem of time series data mining. It aims on finding regular patterns in the time series, usually omitting the exact values and taking only the development or shape of the time series signal. Sometimes this problem is also referred as a rule discovery.\par
One of the common approaches is to have a predefined set of primitive shapes and than to classify the windows of the original sequence into these shapes. The set of primitives can be found by clustering of windows of predefined length. Each sequence is sliced into disjoint windows and than all the windows are clustered into the set of shape primitives. Each primitive is represented by a symbol and than each sequence can be represented by the string of these symbols \cite{Das1998}.\par

\newpage

\section{Subsequence Matching Framework}
As we have outlined in the previous sections, the ubiquity of the time series and its usability in a wide range of applications led the research community to the invention of many various approaches. It was shown that the time series whole matching approach is not sufficient in many areas. Therefore the subsequence matching problem had risen and has been addressed by many researches since the first paper on this topic was published \cite{Faloutsos1994}. Faloutsos et al. have introduced GEMINI framework, an application model for dealing with subsequence matching which in short we can explain in four steps that the application has to perform:
\begin{itemize}
\item slicing the time series sequences into smaller subsequences (originally using sliding window for the indexed data and disjoint window for the query)
\item maping each time series subsequence in a lower dimension. We can call this step segmentation (originally using Fast Fourier Transform)
\item indexing them in multi-dimensional indexing structure (originally R-tree)
\item performing search with a distance function that obeys the rule of lower bounding lemma (originally Euclidean Distance).
\end{itemize}
Despite the fact that the application model is almost two decades old, it is still vital and suits many cases where the subsequence matching procedure has to be employed.
Works that followed that Faloutsos's paper usually tried to enhance only the part of the problem mainly by introducing new methods for dimensionality reduction, new distance functions or a bit of a variance in an indexing and slicing strategy for the subsequence bits. We have mentioned the most significant ones in the previous sections. Those new approaches enhanced the performance of the GEMINI framework but a lot of them were tested only on a constrained, and often artificially created, datasets. It has been shown in \cite{Dinga} that the comparison of those methods is not as clear as some of the authors were trying to declare and so that the results can not be taken as a ground truth. It was argued \cite{Keogh2002} that the experimental results were computed in various environments with various data sets and with various implementations. This problem was previously called \textit{implementation and data bias}.\par
The need for the unified testing dataset was settled and fulfilled by the authors of \cite{Dinga} and a collection of testing datasets has been maintained since then, and the authors of new papers were asked to perform the tests on this collection to bypass at least the \textit{data bias} mentioned above. On the other hand, the \textit{implementation bias} often remains.\par
We would like to address this problem and to present a versatile subsequence matching framework that would be usable for rapid prototyping and testing of the current, but mainly new approaches and would serve as trusted platform for performing these tests. By framework, we mean a complex implementation framework for quick but efficient realization of a wide range of subsequence matching related applications and approaches. It will contain a library of subcomponents, a way to combine them, and an option to create new ones.\par 
We are aware that this goal is not easy one to achieve, because the balance of the boundaries of the framework and the real usability and extensibility must be found so that the developers and scientists would adopt it. Nevertheless, we believe that we can deliver such a framework that can helpful to the community. We can offer the experience with building the MESSIF framework \cite{Batko2007}, part of the MUFIN \cite{Batko2009e} project that allows fast prototyping and testing of general similarity search applications literally on any data domain. Until now, from this paper's point of view, these technologies allowed only dealing with the vectors whole matching problem. What we did was that we used these technologies that were proven by the time and many projects and enriched them with the ability of the subsequence matching.\par
To prove that our solution is really usable for many applications areas, we have decided to demonstrate it on the application related in speech recognition and analysis that covers many problems and many approaches to solve them and where it may not be clear on the first look that the related applications can be solved or enhanced by the state of the art subsequence matching solutions. Later in the text, we will discuss the demo application that demonstrates the combination of state of the art technologies both from the world of speech analysis and subsequence matching.

\subsection{Idea}
\begin{figure}[htbp]
\centering
\includegraphics[scale=0.45]{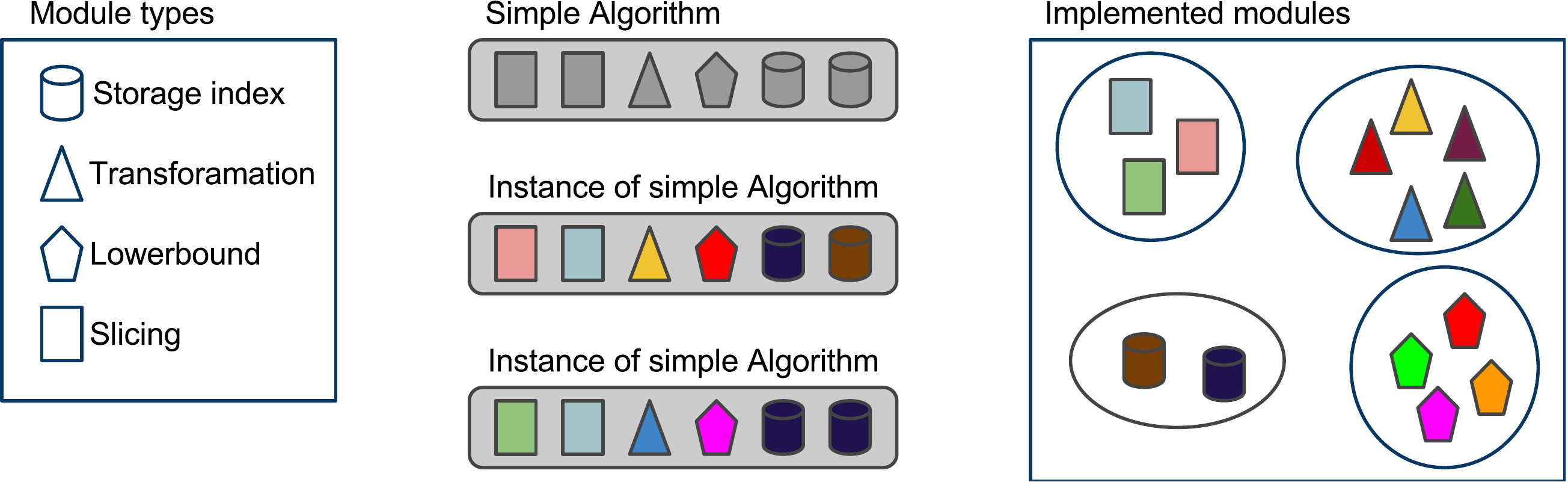}
\caption{SMF principles.}
\label{fig:SMF_modules}
\end{figure}

Our Subsequence Matching Framework relies on the MESSIF library that is designed to help to implement similarity search applications in general. We provide several basic classes and an easy understandable architecture for rapid development and testing of subsequence similarity search applications. We have tried to identify the common sub-problems of the subsequence matching process. We have mentioned some in the rough description of the GEMINI framework but we added some new and still let the architecture opened for implementing brand new approaches. To achieve this we build the framework in a modular way. You can imagine \textit{modules} dealing with:
\begin{itemize}
\item normalization of the time-series/sequence data,
\item transformation for data sequences,
\item storage index for subsequences,
\item slicing of the sequences into the windows,
\item distance function implementations,
\item lower bounding techniques.
\end{itemize}
The concept of \textit{modules} (Fig. \ref{fig:SMF_modules}) should allow good re-usability of the code and will help to overcome  the \textit{implementation bias} because once the module responsible for solving some part of the problem is implemented, it should be used by others.
The architecture allows to implement subroutines with functionality independent of the specific data type. As you can see on the Fig \ref{fig:SMF_modules}, another substantial part of the framework is the definition of algorithms. The algorithms are responsible for the logic of the subsequence matching, indexing and the execution of the queries. Each algorithm declares its parameters. You can imagine that as slots in which the modules are plugged in. So in the example, the Simple algorithm is working with two slicing modules, one transformation module, one distance function that obeys the lower bounding lemma and two indexes -- one for the original sequences and one for the subsequences. Under the definition of the algorithm we can see two instances altering in the usage of modules. This is possible because the implementation of the modules is independent of the data types (to some level of abstraction) and the algorithm itself only uses the interface common to the modules from the same class. These features allow to quickly prototype and alter the algorithm by changing the modules that are used. So when we have the mechanism of the algorithm implemented using such an interface to the the modules declared as a parameters of this algorithm and have implemented a bunch of modules for these classes, we can easily perform lots of combinations and gather the results.

Another great feature of the framework is that altering these combinations does not require to re-compile the code. With the usage of the MESSIF \textit{batch files} one can instantiate new algorithm with new parameters/modules (also instantiated through the \textit{batch file}) and script the whole series of tests in a very comfortable and fast way.\par The binding with the MESSIF framework also brings support for performing various similarity queries including classical range queries, KNN-queries etc.\par
We can also leverage many of state of the art metric indexes previously developed for MESSIF applications like M-index \cite{Novak2009}.

\subsection{Implementation}
As mentioned above, the subsequence matching framework extends the MESSIF platform and is a part of the MUFIN project. The whole MUFIN ecosystem and the SMF are developed using Java and it works with related technologies like Remote Method Invocation (RMI), Java Server Pages (JSP) and use the latest Java features like reflection. We can say that the Subsequence Matching Framework layer extends the MESSIF core layer and that the applications written with aid of the framework can use the MESSIF-UI module for creating the web application demos. The MESSIF offers lots of functionality when developing any similarity search application like indexing and querying in a distributed environment or constructing multi-layered indexing network or composing combined queries.\par
So far we have developed basic skeleton of the subsequence matching framework and now we are trying to proof the concept by implementing various subsequence matching applications with different data domains and different demands. This should help us to find cons of the framework and make the architecture better and more usable for different purposes. The last application domain that has drawn our attention is the speech analysis and generally the audio retrieval.

\newpage

\section{Use Case: Spoken Term Detection}
To prove the applicability of the SMF, we are in a process of implementing an ASR application that would be dealing with the problem of query-by-example spoken term detection. Besides the framework itself, the application will employ one of the state of the art metric indexes. Similarly as in \cite{Hazen2009}, the application will use the indexing based on posteriogram templates that will be extracted with the aid of BUT phonetic recognizer \cite{Schwarz2004}. In contrast with \cite{Hazen2009} we will be using other distance function than DTW, which can be seen as an origin for the performance issues related to the computational demands of DTW and the fact that the DTW itself cannot be indexed as a metric space. We believe that we can achieve very similar result in the terms of quality with the usage of some edit distance based metric function like ERD or ERP which allow better indexing in metric structures like M-index \cite{Novak2009} and that we can achieve better performance results. We would also like to investigate the possibilities for approximate search in the spoken term detection area.\par
\begin{figure}[htbp]
\centering
\includegraphics[scale=0.35]{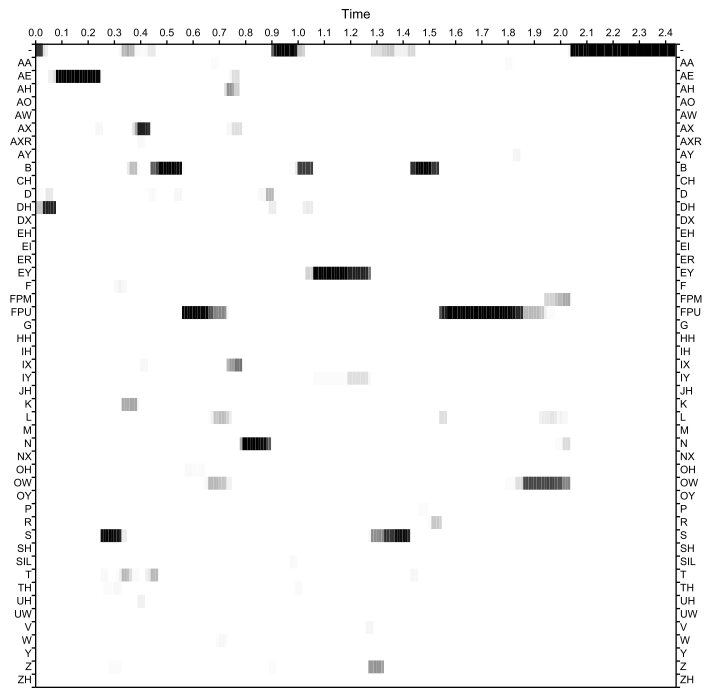}
\caption{An example posteriorgram representation for the spoken phrase ``basketball and baseball'' \cite{Hazen2009}.}
\label{fig:posteriogram}
\end{figure}

The concept of phonetic posteriogram is based on the knowledge of posterior probabilities of every phoneme in the phonetic vocabulary for every time frame. It can be represented as graph of phoneme posterior probability in a time (see Fig. \ref{fig:posteriogram}).
For the evaluation of our approaches we have obtained TIMIT \cite{Garofolo1993} and NIST Spoken Term Detection Development Set \cite{ Group2011} corpora.

\section{Conclusion And Future Work}
We have overviewed basic problems and approaches in audio retrieval area and time-series, or generally sequence, processing. We have outlined that we can apply solutions from subsequence matching area suitable for problems from the audio retrieval area. Our framework should contribute to the solutions in an audio retrieval process by employing subsequence matching techniques. 
In the further research we would like to search for more possibilities for enhancing the performance aspects of the audio retrieval by using metric space approach and advanced subsequence matching techniques. Things like tuning index parameters or deciding on the best subsequence matching strategy for the particular application are yet to be properly investigated.

\section*{Acknowledgments}
This work was supported by national research projects VF20102014004,
MSMT 1M0545, GACR 103/10/0886, and GACR P202/10/P220.

\newpage
\bibliographystyle{plain}
\bibliography{library}
\end{document}